\documentclass[a4paper,12pt]{article}
\usepackage{amssymb,amsfonts,amsmath}

\def\upd{{\rm d}}

\newcommand{\Ro}{{\mathbb R}}

\newcommand{\Abf}{{\bf A}}
\newcommand{\Ebf}{{\bf E}}
\newcommand{\kk}{{\bf k}}
\newcommand{\xx}{{\bf x}}
\newcommand{\yy}{{\bf y}}

\newcommand{\kkph}{{\kk^{\mbox{\tiny ph}}}}
\newcommand{\kkel}{{\kk^{\mbox{\tiny el}}}}

\newcommand{\psiph}{\psi^{\mbox{\tiny ph}}}
\newcommand{\psiel}{\psi^{\mbox{\tiny el}}}
\newcommand{\rhoph}{\rho^{\mbox{\tiny ph}}}
\newcommand{\rhoel}{\rho^{\mbox{\tiny el}}}
\newcommand{\HI}{\hat H^{{\mbox{\tiny I}}}}

\def\ah{\hat a_{\mbox{\tiny H}}}
\def\av{\hat a_{\mbox{\tiny V}}}
\def\ahdagger{\hat a^\dagger_{\mbox{\tiny H}}}
\def\avdagger{\hat a^\dagger_{\mbox{\tiny V}}}

\def\kkph{{\kk^{\mbox{\tiny ph}}}}
\def\kphnull{{\kk^{\mbox{\tiny ph}0}}}
\def\kph{k^{\mbox{\tiny ph}}}

\newcommand{\be}{\begin{eqnarray}}
\newcommand{\ee}{\end{eqnarray}}

\title{On the Emergence of the Coulomb Forces in Quantum Electrodynamics}

\author{J. Naudts\\
\small 
Physics Department, University of Antwerp,\\
\small
Universiteitsplein 1, 2610 Antwerpen, Belgium\\
\small Email: {\tt Jan.Naudts@uantwerpen.be}
}

\begin{document}

\maketitle

\abstract{
A simple transformation of field variables eliminates Coulomb forces from the theory of quantum
electrodynamics. This suggests that Coulomb forces may be an emergent phenomenon
rather than being fundamental. This possibility is investigated in the context
of reducible quantum electrodynamics. It is shown that
states exist which bind free photon and free electron fields.
The binding energy peaks in the long-wavelength limit. 
This makes it plausible that Coulomb forces result from the interaction of the electron/positron
field with long-wavelength transversely polarized photons.
}

\section{Introduction}



Quantum electrodynamics (QED), although very successful, is {\em not}
a mathematically rigorous theory. Several difficulties have to be resolved or circumvented
during the search for a consistent theory.
One of them is the excess number of degrees of freedom in the description of
the electromagnetic fields. In this context the approach of Gupta \cite{GS50} and of Bleuler \cite{BK50}
is well-known. An alternative is found in the work of Creutz \cite{CM79}.
He uses a unitary transformation which yields wave functions which do not
undergo Coulomb forces.  In quantum chromodynamics the light cone gauge is used because then
only transverse gluons remain \cite {BPP98}. In the present work the assumption is that
only transverse photons exist and that the number of degrees of freedom
of the electromagnetic field is two. The argument of Creutz is used 
in opposite direction to define field operators which satisfy the Maxwell equations
in presence of Coulomb forces.

Recently, Verlinde \cite{VEP11,VEP16} developed a cosmological theory in which gravity
forces are emergent forces, in the sense that they are produced by other, more
fundamental, forces of nature. A similar statement is investigated here for the role of Coulomb
forces in QED. Not only are the degrees of freedom of the electromagnetic field
limited to two. In addition the Hamiltonian does not a Coulomb potential.


Further mathematical problems of QED disappear if one abandons the axiom that
representations of the canonical commutation and anti-commutation 
relations must be irreducible. This reducible QED is studied in the work of Czachor and collaborators.
See \cite{CM00,CN06,CW09} and references given in these papers.
This formalism can be simplified along the lines worked out by the author in a series of papers 
\cite{NJ15a,NJ15b,NJ15c,NJ15d}.

The reduction of the representation involves an 
integration over three-dimensional wave vectors. At fixed wave vector the system 
is purely quantum mechanical and consists of a pair of harmonic oscillators at 
each position in space-time to cover transversely polarized photons. A 
16-dimensional Hilbert space describes the different states of an 
electron-positron field.

The mediation of Coulomb forces by transverse photons can be understood
by analogy with the polaron problem of Solid State Physics.
The polaron \cite{DA09} is a state binding an electron with quantized
lattice vibrations. In a polarized medium the electric field of the electron
is completely screened by a redistribution of charges in its vicinity.
The remaining interactions between electron and medium result
in an attractive force between pairs of polarons.
In some situations it is strong enough to form bi-polarons. 

The present work shows that a free electron field can form a bound state
with transversely polarized photons to form a dressed electron field.
By analogy with the polaron case one can then expect that dressed electrons
interact with each other and that Coulomb forces can be explained in this way.

The next section introduces a transformation of field variables which adds
Coulomb forces to transverse photons interacting with a charge field.
The new field variables form what is called here the emergent picture of QED.
They satisfy the full Maxwell equations.
Section \ref {sect:red} highlights some aspects of reducible QED.
Section \ref{sect:bound} discusses the proof of the existence of bound states
of transverse photons in interaction with the electron field.
The final section gives a short discussion of what actually has been achieved.

\section{The emergent picture}

In \cite{CM79} the temporal gauge, also called Weyl gauge or Hamiltonian gauge, is used.
A unitary transformation $\hat V$ is defined by a generator $\hat T(\xx)$
through
\be
\hat V&=&\exp\left(i\int\upd^3 \xx\,\hat T(\xx)\right).
\ee
The generator is of the form
\be
\hat T(\xx)&=&\frac{q}{4\pi}\int\upd^3 \yy\,\hat \Abf(x)\cdot\frac{\xx-\yy}{|\xx-\yy|^3}\hat j_0(\yy).
\ee
Here, $q$ is the elementary unit of charge. Bold characters are used to indicate three-vectors.
The result of \cite{CM79}, in the context of standard QED, is that 
\be
\hat V\nabla\cdot\hat\Ebf\hat V^{-1}&=&\nabla\cdot\hat\Ebf-q\hat j_0,
\ee
where $\hat\Ebf(x)$ are the electric field operators. If they satisfy Gauss's law
in the presence of a charge distribution $\hat j_0(\xx)$
then $\hat V\nabla\cdot\hat\Ebf\hat V^{-1}$ satisfies Gauss's law in absence of charges.

In the present work the trick of \cite{CM79} is directly applied to define new electric
field operators
\be
& &\hat E''_\alpha(x)=\hat E'_\alpha(x)
+\frac{\mu_0 c}{4\pi}\frac{\partial\,}{\partial x^\alpha}\int\upd\yy\,
\frac{1}{|\xx-\yy|}\,\hat U(-x^0)\hat j^0(\yy,0)\hat U(x^0).\cr
& &
\label{emerg:def2}
\ee
Here, $\hat U(x^0)=\exp(-ix^0\hat H/\hbar c)$ is the time evolution of the interacting system.
The new operators are marked with a double prime to distinguish them from the
operators of the non-interacting system and those of the interacting system. The latter are denoted
with a single prime. One verifies immediately that
Gauss' law is satisfied
\be
\sum_\alpha\frac{\partial\,}{\partial x^\alpha}\hat E''_\alpha(x)
&=&
-\mu_0 c\,\hat j^{0\prime}(x).
\ee

The second term of (\ref {emerg:def2}) is the Coulomb contribution to the electric field.
The curl of this term vanishes. Hence it is obvious to take
\be
\hat B''_\alpha(x)&\equiv& \hat B'_\alpha(x).
\ee
This implies the second of the four equations of Maxwell,
stating that the divergence of $\hat B''_\alpha(x)$ vanishes.
Also the fourth equation, absence of magnetic charges,
follows immediately because $\hat E''(x)$ and $\hat E'(x)$
have the same curl. Remains to write Faraday's law as
\be
(\nabla\times\hat B''(x))_\alpha-\frac 1{c}\frac{\partial\,}{\partial x^0}\hat E''_\alpha(x)
&=&
-\mu_0\,\hat j''_\alpha(x)
\ee
with
\be
\hat j''_\alpha(x)
&=&
-\frac 1{\mu_0 c}\frac{\partial\,}{\partial x^0}\left(\hat E''_\alpha(x)-\hat E'_\alpha(x)\right).
\ee
Finally, take $\hat j''_0(x)=\hat j'_0(x)$.
A short calculation shows that the newly defined current operators $\hat j''_\mu(x)$
satisfy the continuity equation.

One concludes that a formalism of QED is possible which does not postulate the
existence of longitudinal or scalar photons. Two pictures coexist: the original
Heisenberg picture and what is called here the {\em emergent} picture.
In both pictures the time evolution of all operators is the same, but
the definition of the electromagnetic field operators differs.
In the original description only transversely polarized photons exist.
On the other hand, the field operators of the emergent
picture satisfy the full Maxwell equations, including Coulomb forces.

\section{Reducible QED}
\label{sect:red}

A dominant characteristic of reducible QED, in the version used here, is that
many expressions look familiar from standard QED, except that integrations over the
wave vector are missing. They are postponed to the evaluation of quantum expectation values.
As a consequence, field operators depend on both
position $x$ in spacetime and wave vector $\kkph$ in $\Ro^3$.
For instance, the electromagnetic potential operators are defined by
\be
\hat A_\alpha(x)&=&\frac 1{2}\lambda\varepsilon^{(H)}_\alpha(\kkph)
\left[e^{-i\kph_\mu x^\mu}\ah+e^{i\kph_\mu x^\mu}\ahdagger\right]\cr
& &+\frac 1{2}\lambda\varepsilon^{(V)}_\alpha(\kkph)
\left[e^{-i\kph_\mu x^\mu}\av+e^{i\kph_\mu x^\mu}\avdagger\right].
\label{em:potop}
\ee
Here, $\ah,\ahdagger,\av,\avdagger$ are the creation and annihilation operators of
horizontally, respectively vertically polarized photons.
The dispersion relation of the photon is $\kphnull=\hbar|\kkph|$ as usual.
The  $\varepsilon^{(H)}_\alpha(\kkph)$ and $\varepsilon^{(V)}_\alpha(\kkph)$
are polarization vectors. The parameter $\lambda$ is introduced for dimensional reasons.

A consequence of the missing integration over the wave vector is that
equal-time fields become non-commutative.
For instance, a calculation starting from (\ref {em:potop}) shows that
\be
\left[\hat A_\alpha(\xx,0),\hat A_\beta(\yy,0)\right]_-&=&
\frac{i\lambda^2}{2} F_{\alpha,\beta}(\kkph)
\sin(\kkph\cdot(\xx-\yy)),
\ee
where $F(\kkph)$ projects onto the plane orthogonal to $\kkph$.
If the integration over $\kkph$ is executed then the standard result of
vanishing equal time commutators follows.

Another feature of the theory is that wave functions are properly normalized for each
wave vector separately. For instance, if $\psi$ describes an electron/positron field then
$\psi_\kk$ is a wave function in the $\kk$-th Fock space and satisfies 
$\langle\psi_\kk|\psi_\kk\rangle=1$ for each value of $\kk$.
Superpositions of wave functions with different wave vector are allowed. The general
wave function is therefore of the form
\be
\psi_{\kk}&=&\sum_{X\subset\{1,2,3,4\}}\sqrt{\rho_X(\kk)}\,|X\rangle\quad\mbox{ for any }\kk.
\ee
The set $X$ selects one of the 16 possible states of an electron/positron field.
The empty set $\emptyset$ refers to the vacuum state $|\emptyset\rangle$.
Normalization requires that
\be
\sum_{X\subset\{1,2,3,4\}}\rho_X(\kk)=1\quad\mbox{ for any }\kk.
\ee
Ultraviolet divergences are avoided by requiring that $\rho_X(\kk)$ with $X\not=\emptyset$ vanishes for
large values of $|\kk|$.
Similarly, the general wave function of the free electromagnetic field is of the form
\be
\psi_{\kkph}&=&\sum_{m,n=0}^\infty\sqrt{\rho_{m,n}(\kkph)}e^{i\phi(m,n)}|m,n\rangle.
\ee
Normalization requires that
\be
\sum_{m,n=0}^\infty\rho_{m,n}(\kkph)&=&1\quad\mbox{ for any }\kkph.
\ee

The Dirac currents $\hat j^\mu(x)$ are defined in terms of Dirac spinors $\hat\psi_r(x)$
which satisfy a free Dirac equation. See \cite{NJ15b} for details.
The Dirac equation is only used to define currents $\hat j^\mu(x)$ in absence of interaction
with the electromagnetic fields. No interacting Dirac equation is considered.
Instead, the interactions between the free field operators are described
by the usual interaction Hamiltonian in a Heisenberg picture.
See (\ref {bound:HI}) below.

\section{Bound states}
\label{sect:bound}

Let $\hat b^\dagger_{\uparrow}$ denote the creation operator for an electron with spin up.
An example of a realistic electron field is described by
\be
\psiel&=&e^{i\chi(\kk)}\sqrt{1-\rhoel(\kk)}|\emptyset\rangle
+\sqrt{\rhoel(\kk)}\hat b^\dagger_{\uparrow}|\emptyset\rangle.
\ee
Similarly, a realistic wave function for a horizontally polarized photon is
\be
\psiph &=&\sqrt{\rhoph(\kkph)}\ahdagger|\emptyset\rangle
+\sqrt{1-\rhoph(\kkph)}|\emptyset\rangle.\cr
& &
\ee

The Hamiltonian is the usual one, with interaction part
\be
\HI&=&\int\upd \xx\,\hat j^\mu(\xx,0) \hat A_\mu(\xx,0).
\label{bound:HI}
\ee
Assume $\rhoel(-\kkel)=\rhoel(\kkel)$.
Then the average interaction energy of a product state
$\psi=\psiph\psiel$ vanishes for symmetry reasons.
However, there exist entangled wave functions
which lower the total energy.

Choose for instance an entangled wave function of the form
\be
\psi_{\kkph,\kk}
&=&
[\tau(\kkph,\kk)\ahdagger+1-\tau(\kkph,\kk)]\,
\sqrt{\rho(\kkph)\rhoel(\kk)}\,\hat b^\dagger_{\uparrow}|\emptyset\rangle,\cr
& &
+\sqrt{1-\rho(\kkph)\rhoel(\kk)}|\emptyset\rangle,
\ee
where $\tau(\kkph,\kk)$ equals either 1 or 0.
Assume in addition that the electron density
$\rhoel(\kk)$ has a Gaussian shape. Then an explicit calculation shows that
the binding energy peaks for long wavelength horizontally polarized 
photons with wave vector $\kkph$ in principle direction 1.
See \cite{NJ15c} for detailed calculations.
Moreover the total energy is lower than that of the free electron.
One concludes that there exists states binding an electron field and
a transversely polarized photon field.

\section{Discussion}

In a theory with only transverse photons and no Coulomb forces a simple transformation
of the field variables, given by (\ref {emerg:def2}), introduces new variables
which satisfy the full Maxwell equations. The new variables form what is called here the emergent
picture of QED. 
The time evolution of operators is the same in the emergent picture as in the original Heisenberg
picture. Hence one can avoid to introduce Coulomb forces if one does not want to have them.

Is this transformation more than a mathematical trick?
A plausible explanation of the physics behind this transformation 
is that long-wavelength transverse photons produce effective forces between
different parts of the electron/positron field. The expectation is that
these effective forces coincide with what is known as Coulomb forces.

Investigation of the scenario sketched above starts with
a mathematical proof that transverse photons do interact with the electron/ positron
field and even can form bound states. This proof is given in the context
of reducible QED because this formalism allows for a mathematically rigorous
treatment. Details of the proof are found elsewhere \cite{NJ15c}.
The next thing to do is an analysis of the time evolution of these bound states.
This analysis is still missing.

\bigskip
The author declares that there is no conflict of interest regarding the publication of this paper.

\end{document}